\documentclass[twoside,slac_one]{revtex4}
\usepackage{graphicx}
\usepackage{fancyhdr}
\usepackage{amsmath} % American Mathematics Society standards
\usepackage{bm}% bold math
\usepackage{amsxtra}
\usepackage{amssymb}
\usepackage{amsthm}
\usepackage{latexsym}
\usepackage{lscape}

\begin{document}

\title{A Search for Higgs Boson in $H\rightarrow W^+W^-$}


\author{Kevin Sung on behalf of the CMS Collaboration}
\affiliation{Massachusetts Instititute of Technology, Cambridge, MA, USA}

\begin{abstract}
A search for the Higgs boson decaying to $W^+W^-$ has been performed on $1.1\:$fb$^{-1}$ of pp 
collision data at $\sqrt{s}=7\:$TeV collected with the Compact Muon Solenoid (CMS) detector in 2011.
No significant excess above Standard Model background expectation is observed, and upper limits on
Higgs boson cross section production are derived, excluding the presence of a Higgs boson with mass
in the range of $\left[150,\,193\right]\:$GeV$/c^{2}$ at $95\%$ confidence level.
\end{abstract}

\maketitle

\thispagestyle{fancy}
\section{Introduction}
The Standard Model (SM) of particle physics successfully explains the majority of high-energy 
experimental data~\cite{ref1}. One of the key remaining questions is the origin of masses of 
elementary particles. In the SM, it is attributed to spontaneous breaking of the electroweak 
symmetry~\cite{ref2,ref3,ref4}. The existence of the field quantum, the Higgs boson, has still to be
experimentally confirmed.

The first search for the Higgs boson at the LHC was published by CMS based on a data sample recorded
in 2010 using the fully leptonic final state~\cite{ref8}. The strategy adopted in~\cite{ref8} has 
been improved to cope with a larger number of interactions within a single bunch crossing, referred
to as \emph{pile-up}. Furthermore, to improve signal sensitivity, lower transverse momentum ($p_T$)
and events with one and two reconstructed jets are now considered. The analysis is performed using 
an integrated luminosity of $1.1\pm0.1\:$fb$^{-1}$~\cite{hww_pas_eps}.

The analysis separates the full set of $H\rightarrow WW\rightarrow \ell\nu\ell'\nu'$ events into 
three categories according to the event jet multiplicity: $H+0\:$jets, $H+1\:$jet, and $H+2\:$jets.
$W^+W^-$ candidates with both $W$ bosons decaying leptonically are selected in final states 
consisting of two isolated, high $p_T$, oppositely charged leptons (electrons or muons) and large 
missing transverse energy due to the undetected neutrinos. The search for the Higgs boson is 
performed in the mass range of $115-600\:$GeV$/c^{2}$, using both a cut-based signal extraction and
a multivariate analysis. All Higgs production mechanisms are considered as part of the signal: the 
gluon fusion process, a Higgs boson in the final state accompanied by a $W$ or $Z$ boson or by a 
pair of top quarks, and the vector boson fusion (VBF) process. The latter process plays a major
role in the $2$-jet category. The expected production cross sections for a SM Higgs boson are taken
from~\cite{ref9}.

\section{Event Preselection}
Several SM processes can lead to a final state similar to that of $H\rightarrow W^+W^-$ signal, in 
addition to the non-resonant $W^+W^-$ process. These backgrounds include: $W$+jets and QCD multi-jet
events where at least one of the jets is misidentified as a lepton, top production ($t\bar{t}$ and 
$Wt$), the Drell-Yan $Z/\gamma^*\rightarrow\ell^+\ell^-$ process, and diboson production 
($W\gamma$, $WZ$, $ZZ$).

Events are selected with two oppositely charged, isolated leptons with high transverse momenta 
($p_T$), in three final states: $e^+e^-$, $\mu^+\mu^-$, and $e^\pm\mu^\mp$. These final states 
include $W\rightarrow\tau\nu_{\tau}$ events with leptonic $\tau$ decays. The trigger system requires 
the presence of one or two high $p_T$ leptons. The trigger efficiency for signal events is found 
to be above $97\%$ in the $\mu^+\mu^-$ final state and above $99\%$ in the $e^+e^-$ and 
$e^\pm\mu^\mp$ final states for a Higgs boson with $m_H \sim 160\:$GeV$/c^2$.

Charged leptons from $W$ boson decays are expected to be isolated from other activity in the event. 
For each lepton candidate, a cone of radius 
$\Delta R \equiv \sqrt{\Delta\eta^2 + \Delta\phi^2} < 0.3\,(0.4)$ for muons (electrons) is 
constructed around the track direction at the event vertex. The scalar sum of the transverse energy
of charged particle candidates~\cite{ref20} compatible with the chosen primary vertex and contained 
within the cone is calculated, excluding the contribution from the lepton candidate itself. If this 
sum exceeds some fraction of the candidate $p_T$, the lepton is rejected. The exact requirement 
depends on lepton $\eta$, $p_T$, and flavor.

Neutrinos from $W$ boson decays escape detection. This results in an imbalance in the measured 
energy depositions in the transverse plane, denoted by $E^{\mbox{\scriptsize{miss}}}_T$. To reject 
$Z/\gamma^*\rightarrow\tau^+\tau^-$ background events as well as $Z/\gamma^*\rightarrow\ell^+\ell^-$
events with mis-measured $E^{\mbox{\scriptsize{miss}}}_T$ associated with poorly reconstructed 
leptons, we use the \emph{projected} $E^{\mbox{\scriptsize{miss}}}_T$~\cite{ref21}. This is 
calculated as the component of $E^{\mbox{\scriptsize{miss}}}_T$ transverse to the closest lepton if
it is closer than $\pi/2$ in azimuthal angle, and the full $E^{\mbox{\scriptsize{miss}}}_T$ 
otherwise. To reduce the dependence of the \emph{projected} $E^{\mbox{\scriptsize{miss}}}_T$ on 
pile-up we use the minimum of two different estimators: the first includes all particle candidates 
in the event~\cite{ref22}, while the second uses only the charged particle candidates associated to 
the primary vertex. This exploits the correlation between the two estimators in events with real 
\emph{projected} $E^{\mbox{\scriptsize{miss}}}_T$, as in the signal, and uncorrelation otherwise, 
as in Drell-Yan events. Events are required to have \emph{projected} 
$E^{\mbox{\scriptsize{miss}}}_T$ above $40\:$GeV in the $e^+e^-$ and $\mu^+\mu^-$ final states, and 
above $20\:$GeV for the $e^\pm\mu^\mp$ final states, which have lower contamination from 
$Z/\gamma^*\rightarrow\ell^+\ell^-$ decays. To reduce the contamination from 
$Z/\gamma^*\rightarrow\ell^+\ell^-$ events where the $Z$ boson recoils against a jet, the angle in 
the transverse plane between the dilepton system and the most energetic jet must be smaller than 
$165\:$degrees in the $ee/\mu\mu$ final states. This selection is only applied if the leading jet 
$p_T>15\:$GeV$/c$. 

To further reduce the Drell-Yan background in the $ee$ and $\mu\mu$ final states, events with a 
dilepton invariant mass within $\pm\:15\:$GeV$/c^2$ of the $Z$ mass are rejected. Events with 
dilepton masses below $12\:$GeV$/c^2$ are also rejected to suppress contributions from low mass 
resonances.

Jets are reconstructed from calorimeter and tracker information using the anti-$k_T$ clustering 
algorithm~\cite{ref24} with distance parameter $R=0.5$. Jets are required to have $p_T>30\:$GeV$/c$
within $|\eta|<5.0$. To correct for the pile-up contribution to the jet energy, a method to 
determine a median energy density per event is applied~\cite{ref25} to compute the offset to be 
subtracted from the jet $p_T$~\cite{ref26}.

To suppress the top quark background, we apply a \emph{top veto} based on soft-muon and $b$-jet 
tagging~\cite{ref27,ref28}. The first method vetoes events containing muons from the $b$-quark 
decays. The second method uses standard $b$-jet tagging looking for tracks with large impact 
parameter within jets. The algorithm is applied also in the case of zero counted jets, which can 
still contain low $p_T$ tagged jets.

To reduce the background from diboson processes, such as $WZ$ and $ZZ$ production, any event that 
has an additional third lepton passing the identification and isolation requirements is rejected. 
$W\gamma$ production, where the photon is misidentified as an electron, is suppressed by stringent 
$\gamma$ conversion rejection requirements. After applying all preselection requirements, $626$, 
$334$, $175$ events are observed in data coming from $0$-jet, $1$-jet, and $2$-jet categories, 
respectively. This sample is dominated by $W^+W^-$ events.

\section{Higgs Signal Extraction}
To enhance the sensitivity to the Higgs signal, two different analyses are performed in the $0$-jet 
and $1$-jet categories. The first analysis is a cut-based approach where further requirements on a 
few observables are applied, while the second analysis makes use of multivariate techniques. Both of
them cover a large Higgs boson mass range, and each is separately optimized for different $m_H$ 
hypotheses. The first method is the simplest approach to be performed on the recorded data sample. 
The second method is more powerful, since it exploits the information present in the correlation 
among the variables. The $2$-jet category makes use of a simple cut-based approach for now, since 
in any case the sensitivity with the current available integrated luminosity is limited.

\subsection{Strategy in the 0-jet and 1-jet Categories}
\begin{figure}[htb]
\centering
\includegraphics[width=80mm]{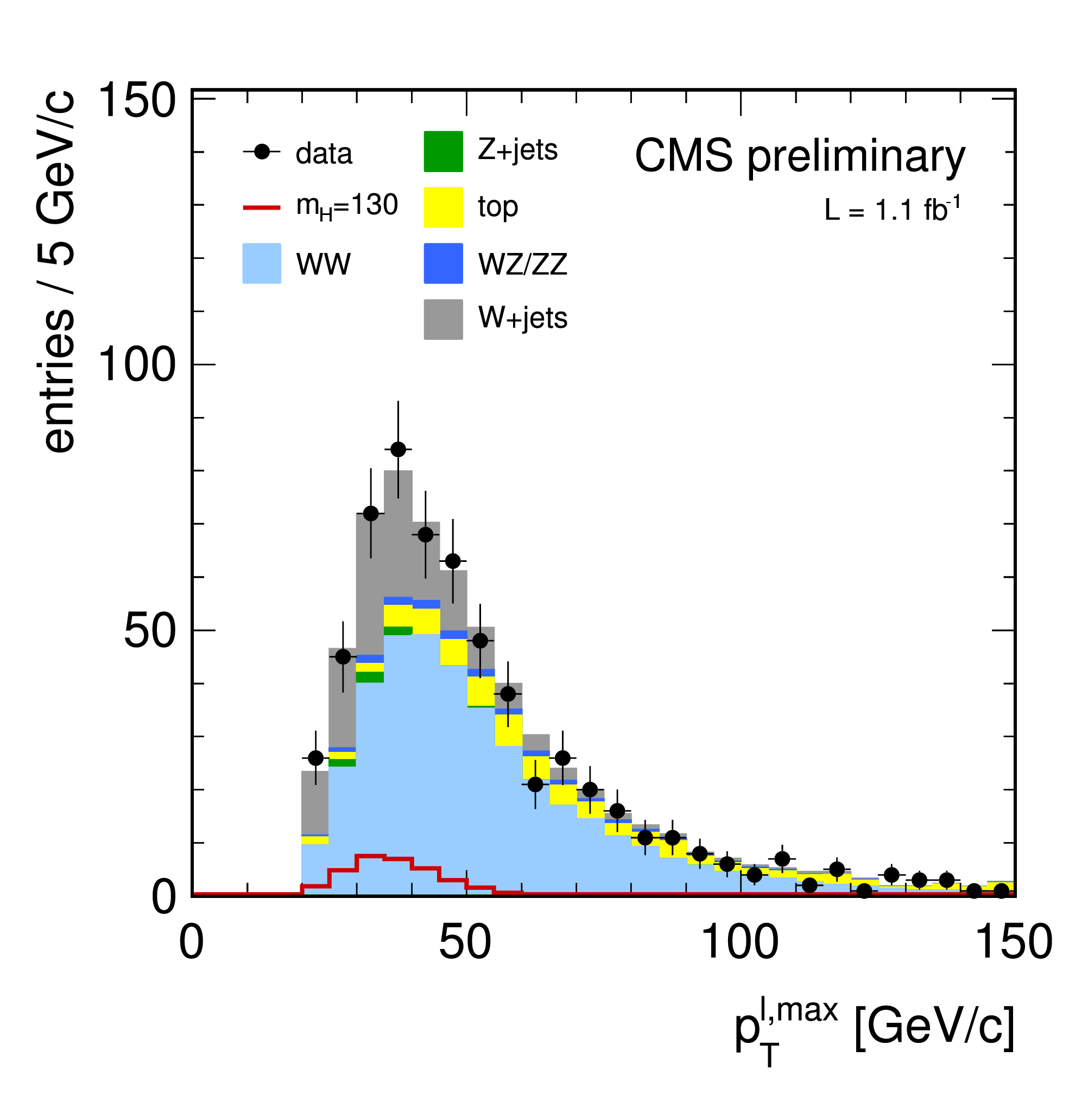}
\includegraphics[width=80mm]{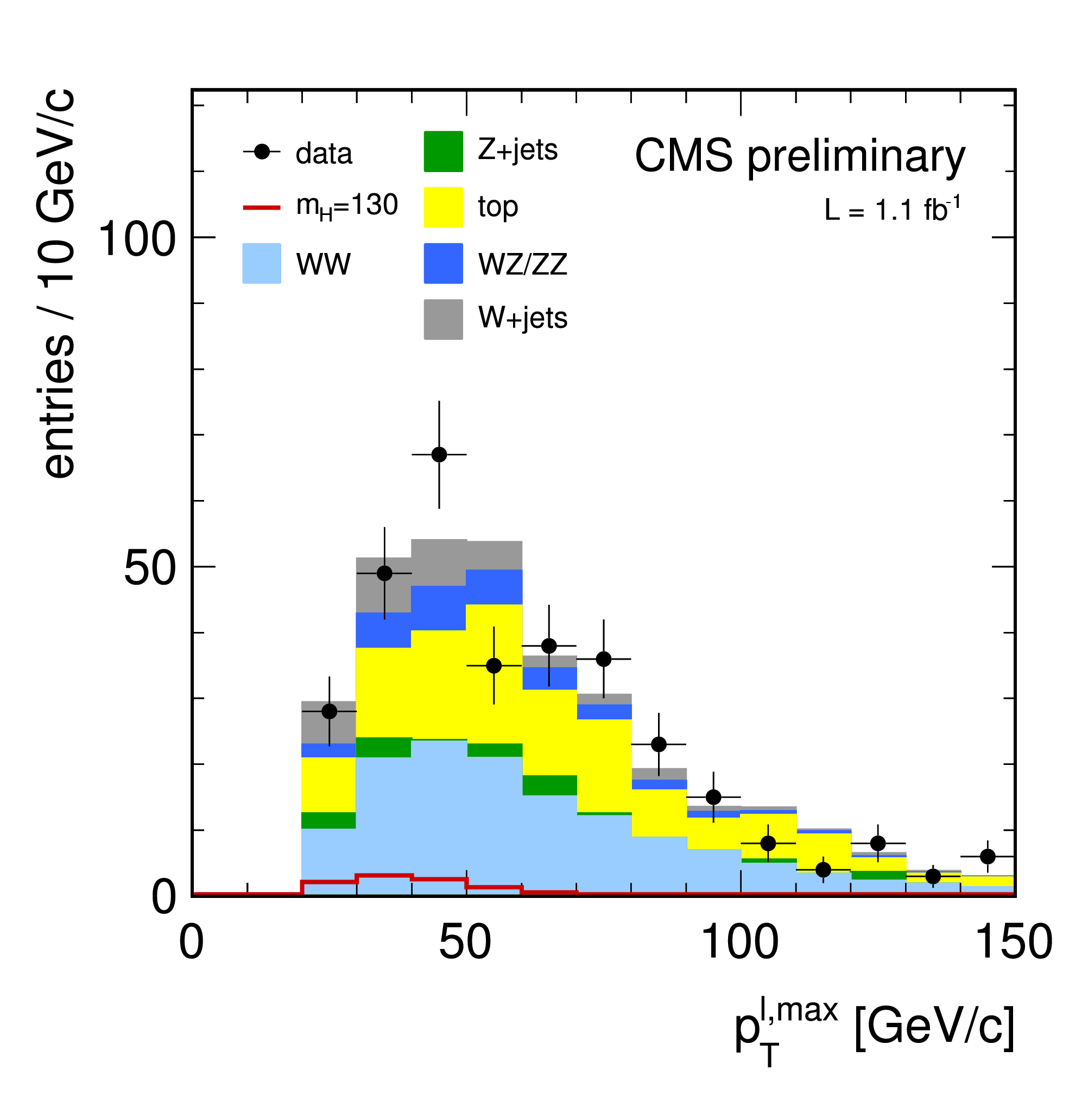}
\caption{Transverse momentum distributions of the harder lepton for the $0$-jet (left) and $1$-jet 
(right) categories.}
\label{fig:ptmax_0j_1j}
\end{figure}

\begin{figure}[htb]
\centering
\includegraphics[width=80mm]{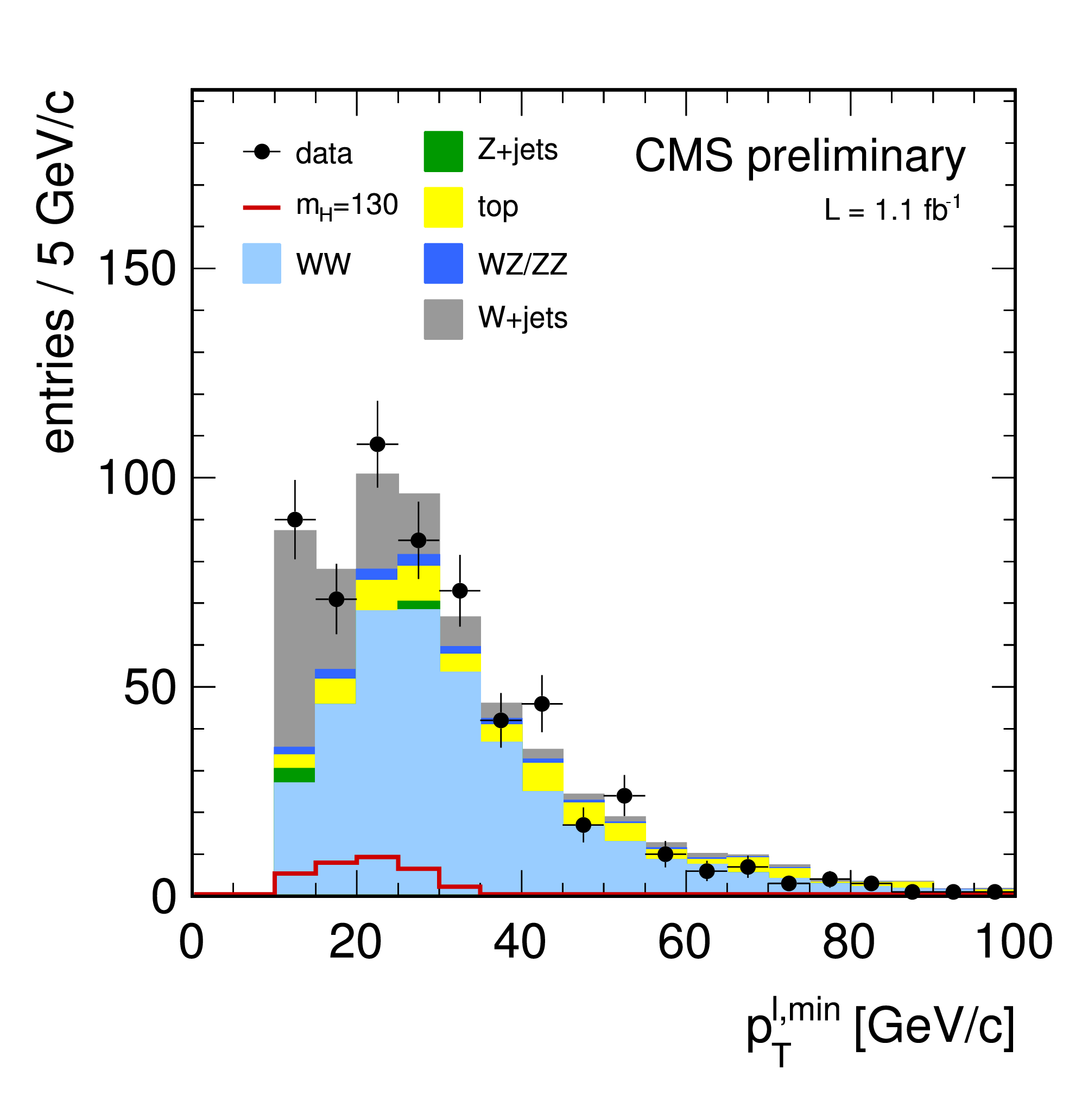}
\includegraphics[width=80mm]{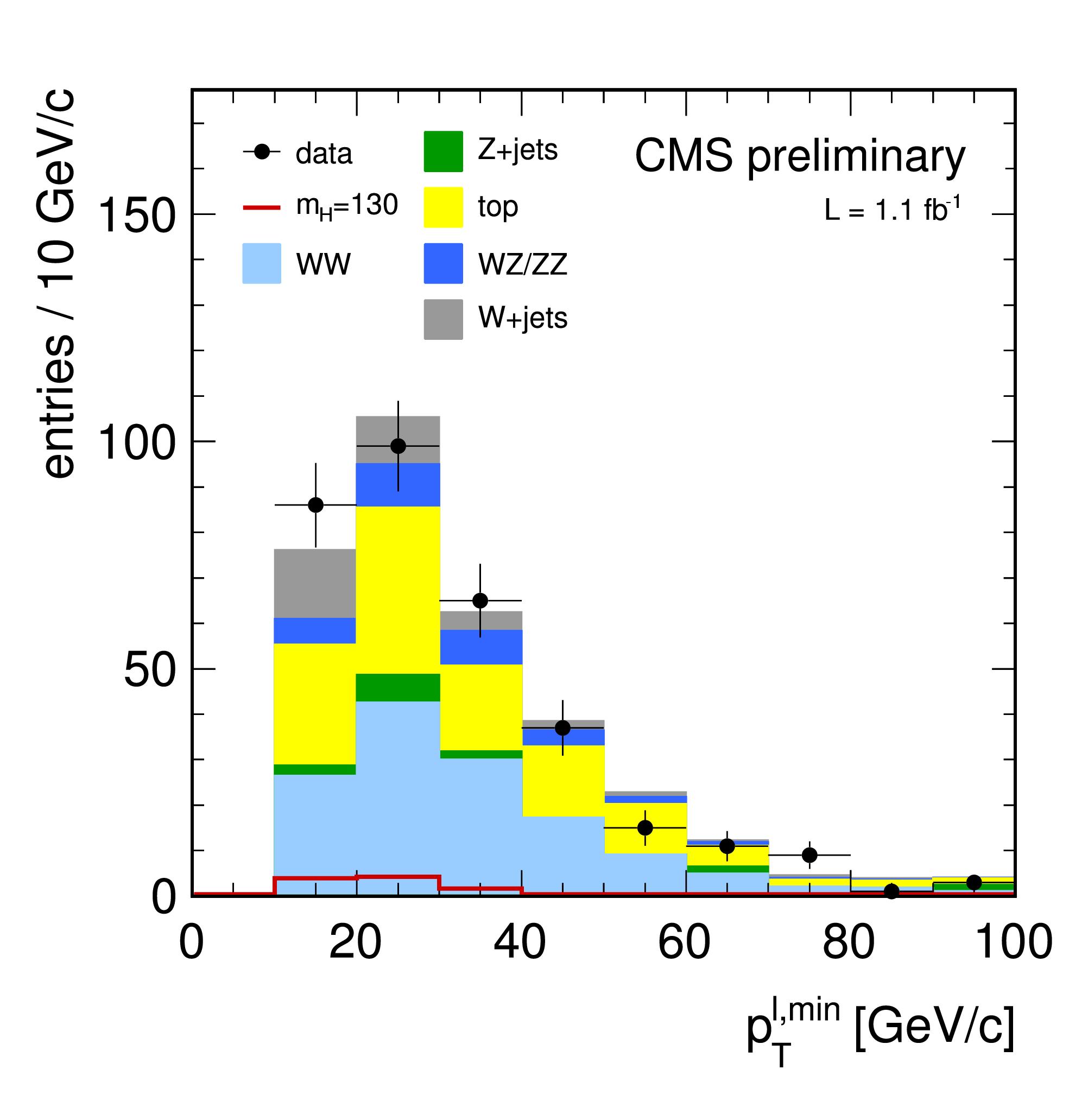}
\caption{Transverse momentum distributions of the softer lepton for the $0$-jet (left) and $1$-jet 
(right) categories.}
\label{fig:ptmin_0j_1j}
\end{figure}

\begin{figure}[htb]
\centering
\includegraphics[width=80mm]{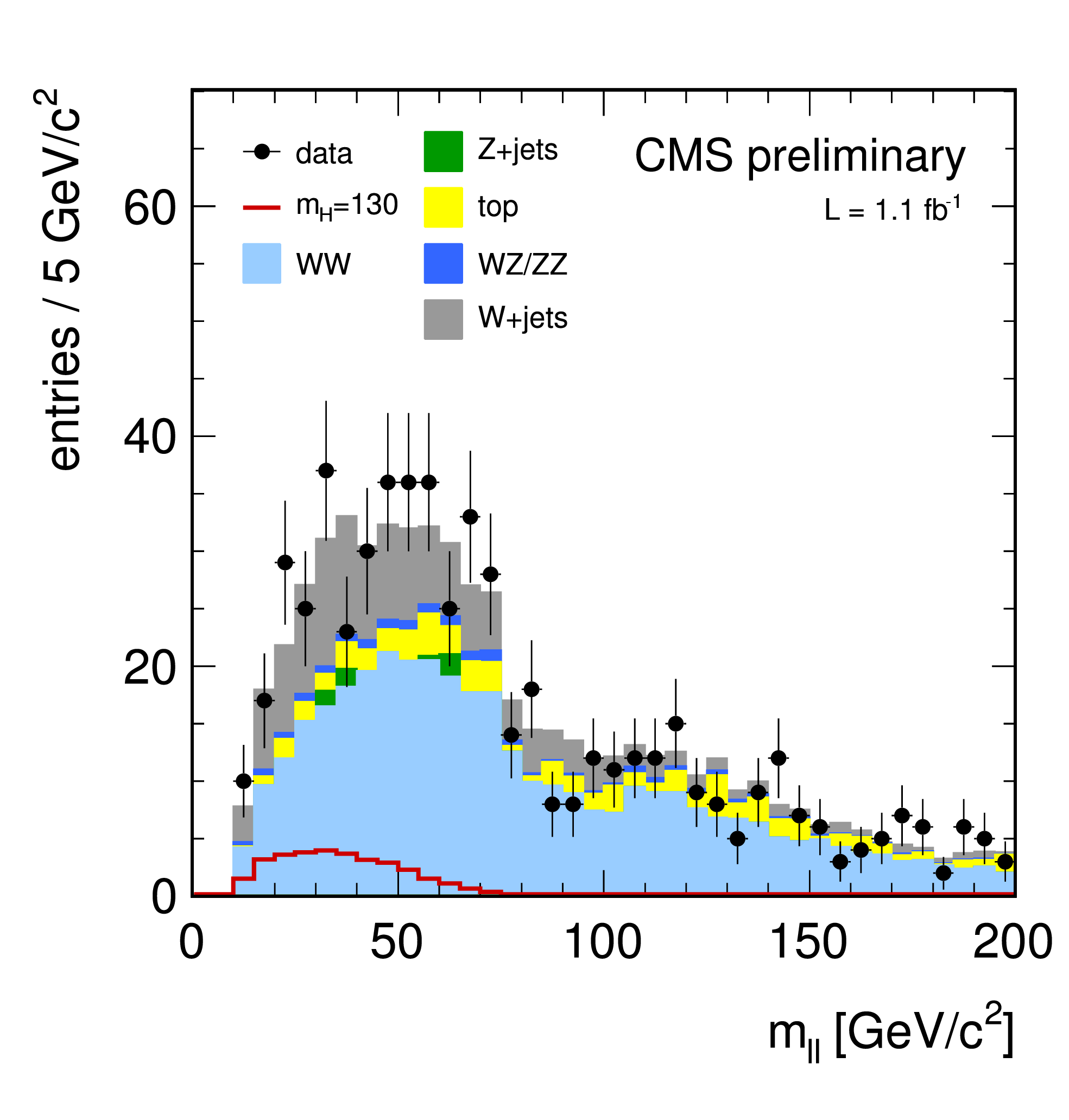}
\includegraphics[width=80mm]{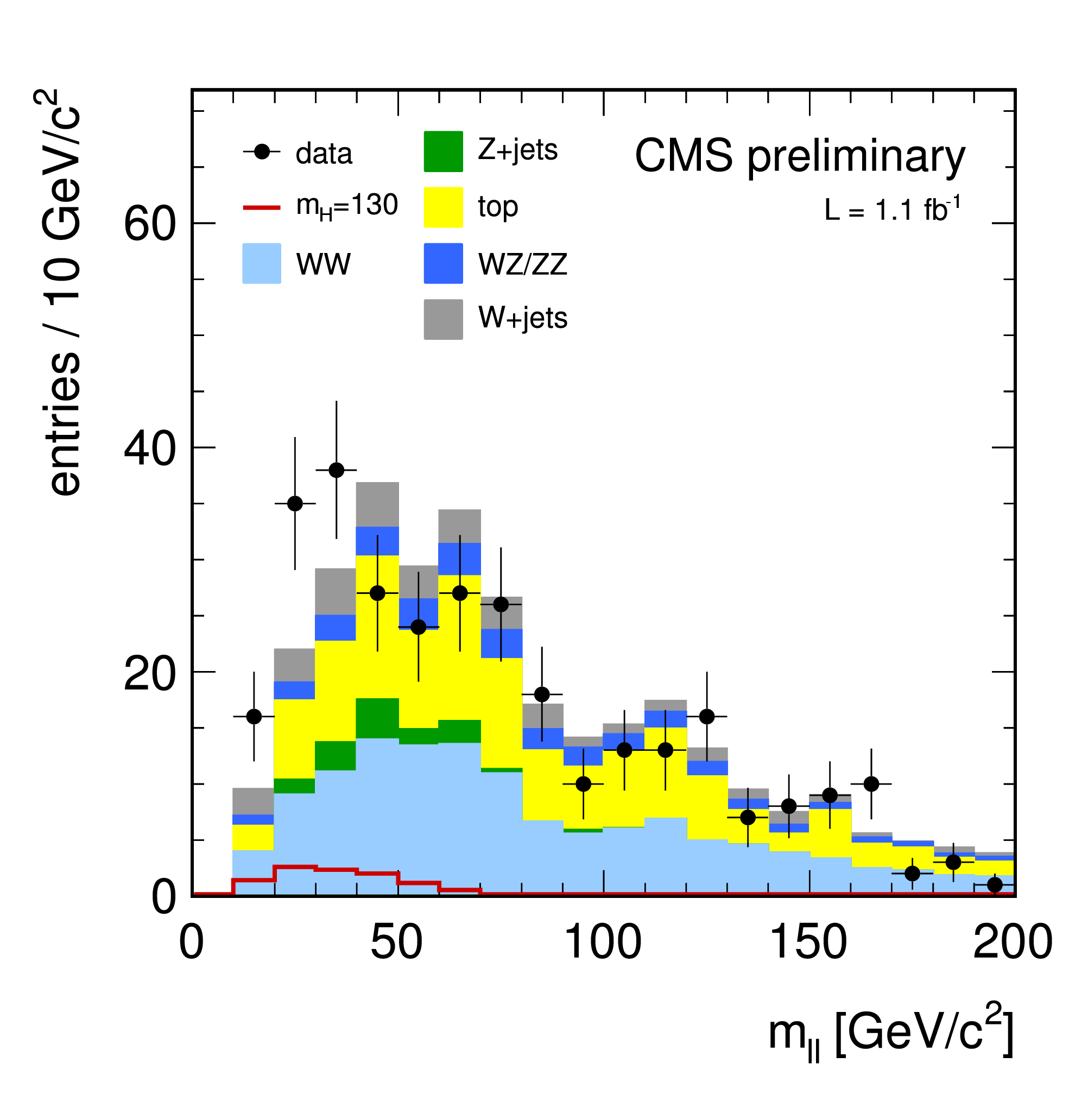}
\caption{Dilepton mass distributions for the $0$-jet (left) and $1$-jet (right) categories.}
\label{fig:dilmass_0j_1j}
\end{figure}

\begin{figure}[htb]
\centering
\includegraphics[width=80mm]{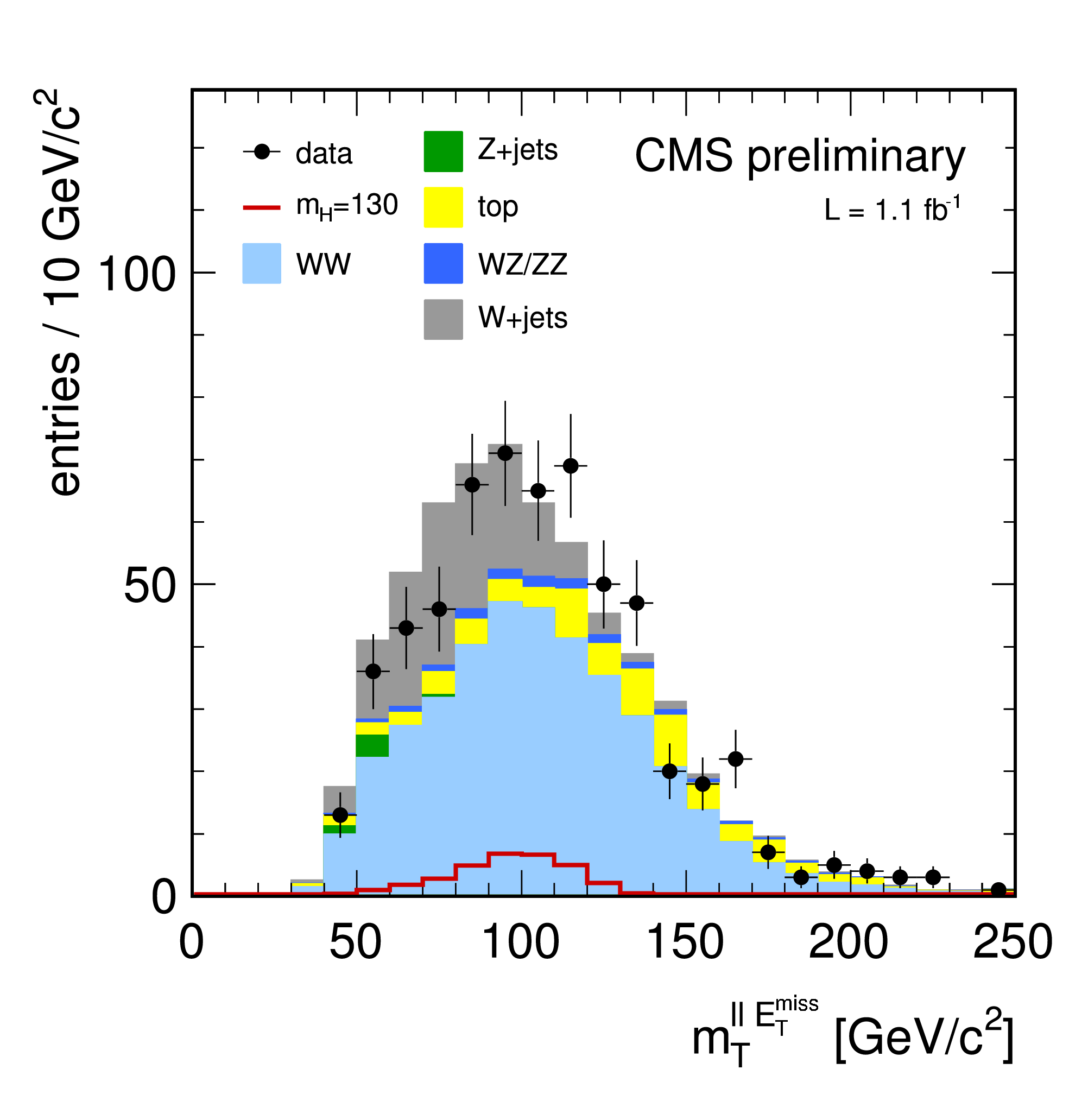}
\includegraphics[width=80mm]{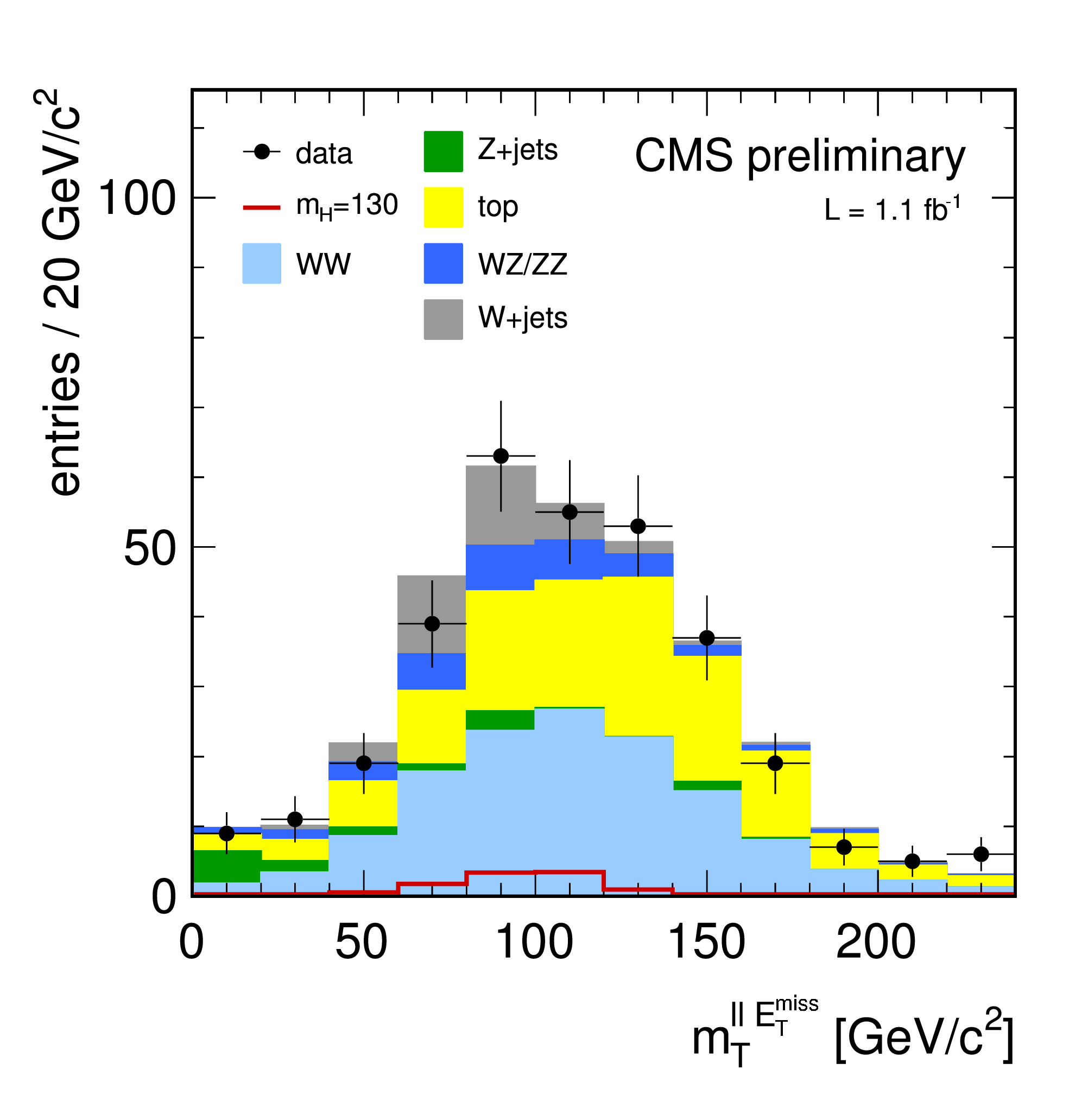}
\caption{Transverse mass distribution for the $0$-jet (left) and $1$-jet (right) categories.}
\label{fig:mt_0j_1j}
\end{figure}

\begin{figure}[htb]
\centering
\includegraphics[width=80mm]{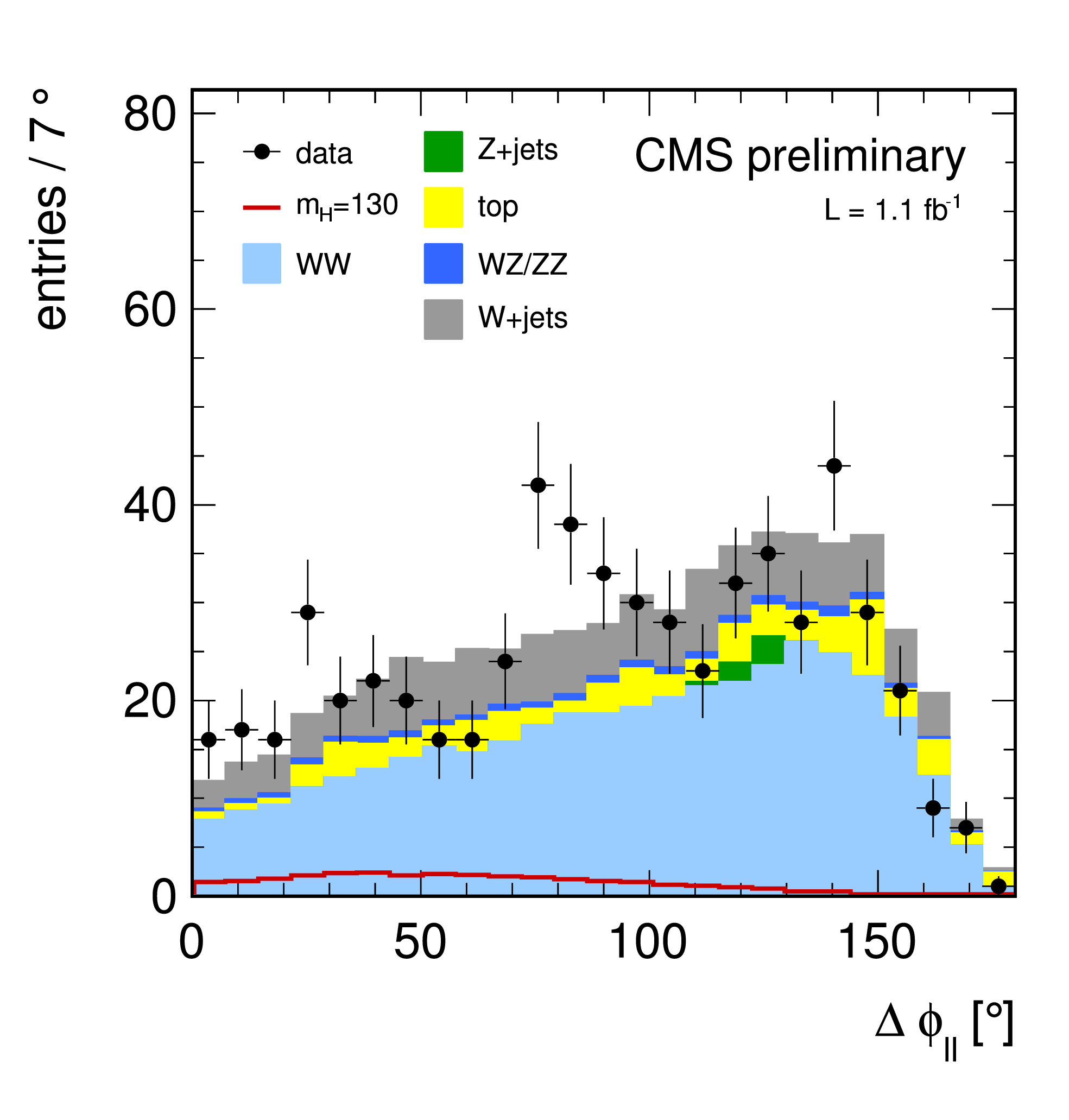}
\includegraphics[width=80mm]{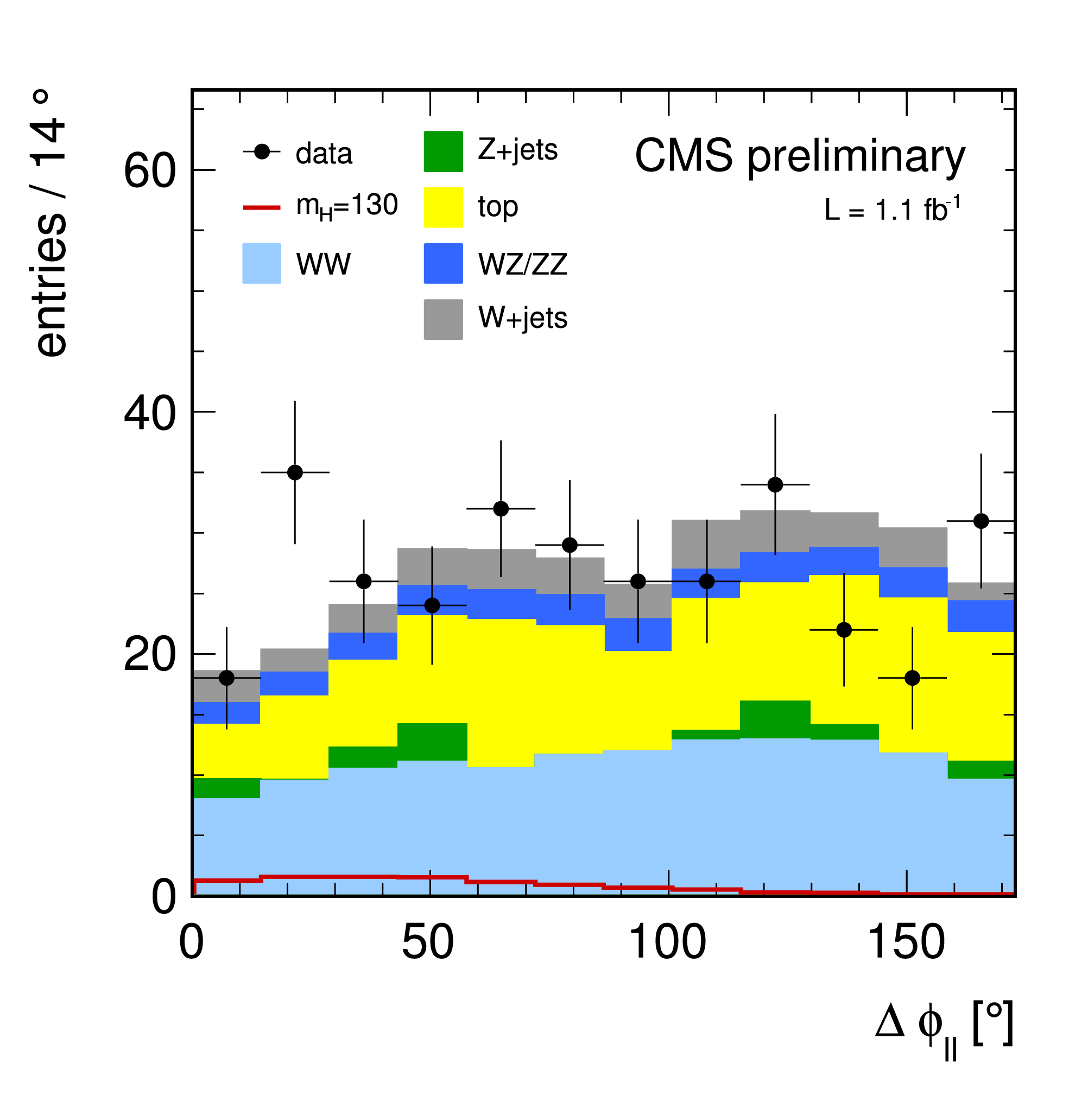}
\caption{Azimuthal angle difference between the two selected leptons for the $0$-jet (left) and 
$1$-jet (right) categories.}
\label{fig:deltaphi_0j_1j}
\end{figure}

In the cut-based approach, extra requirements are placed on the transverse momenta of the harder 
($p^{\mbox{\scriptsize{max}}}_T$) and softer ($p^{\mbox{\scriptsize{min}}}_T$) leptons, the dilepton
mass ($m_{\ell\ell}$), the transverse Higgs mass ($m^{\ell\ell\,E_T^{\mbox{\scriptsize{miss}}}}_T$), 
and the azimuthal angle difference between the two selected leptons ($\Delta\phi_{\ell\ell}$). 
Figures~\ref{fig:ptmax_0j_1j} - \ref{fig:deltaphi_0j_1j} show these distributions after 
preselection, for a SM Higgs signal with $m_H=130\:$GeV$/c^2$ and for backgrounds, in the $0$-jet 
and $1$-jet categories. 

\begin{figure}[htb]
\centering
\includegraphics[width=80mm]{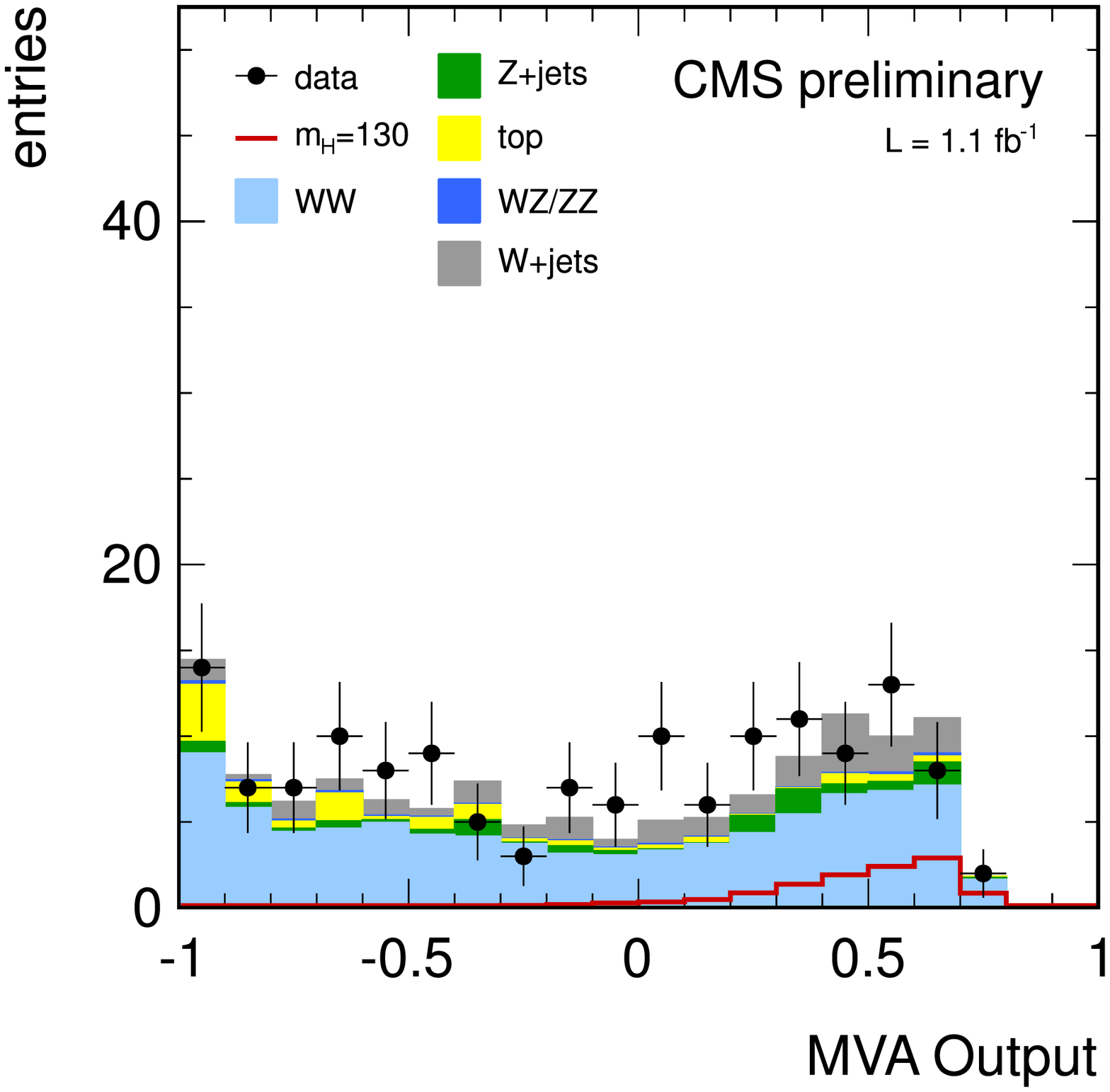}
\includegraphics[width=80mm]{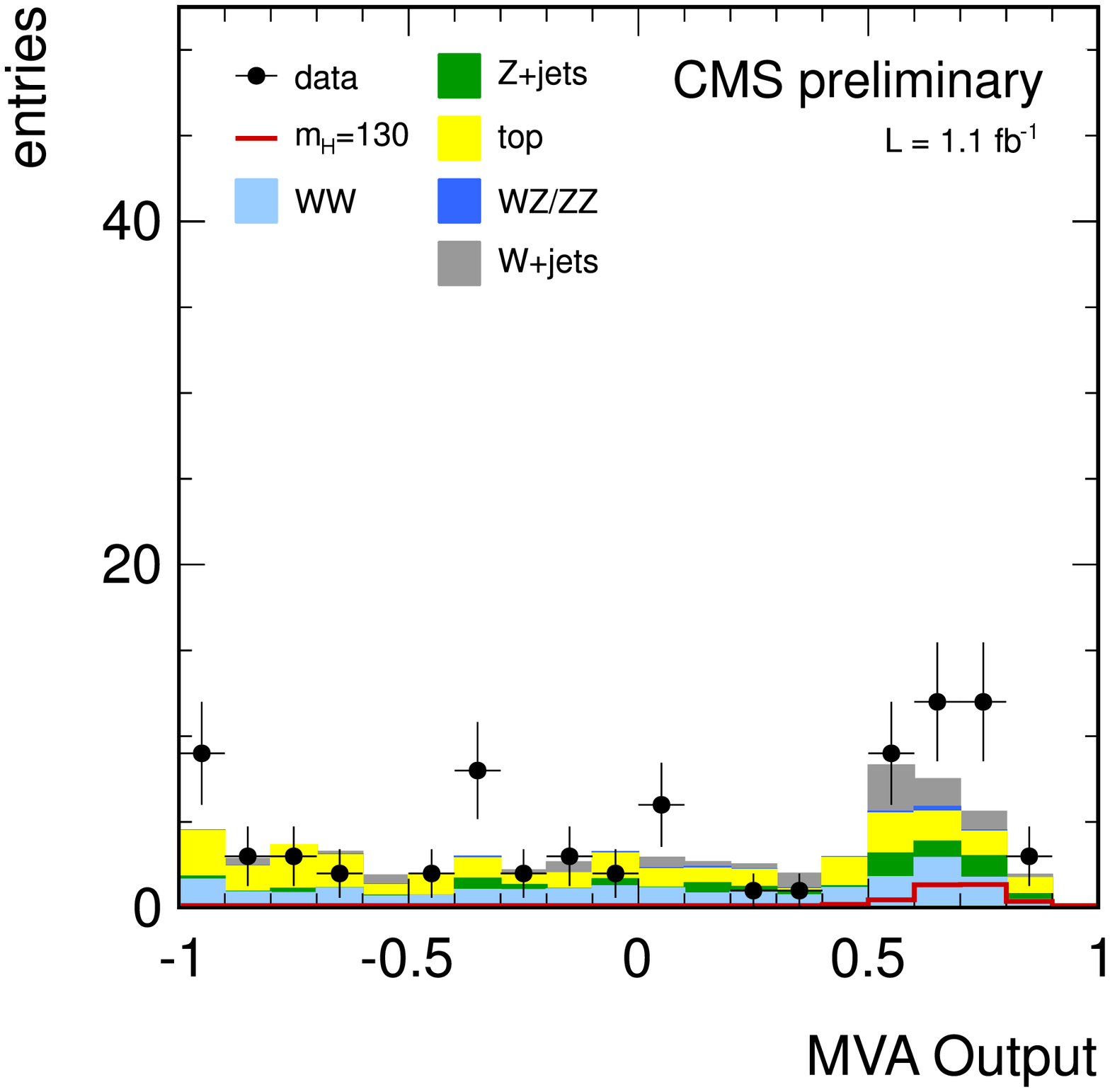}
\includegraphics[width=80mm]{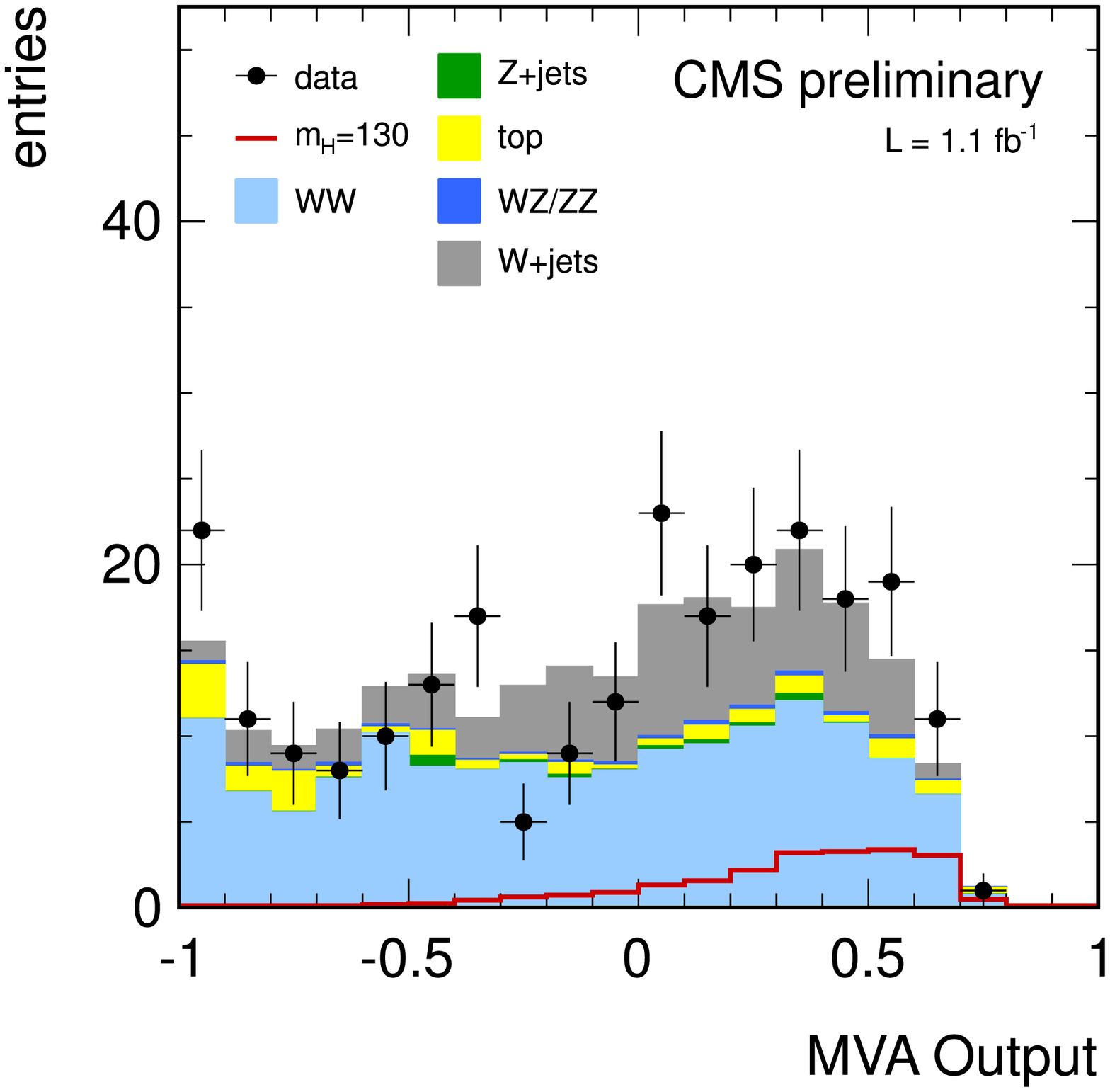}
\includegraphics[width=80mm]{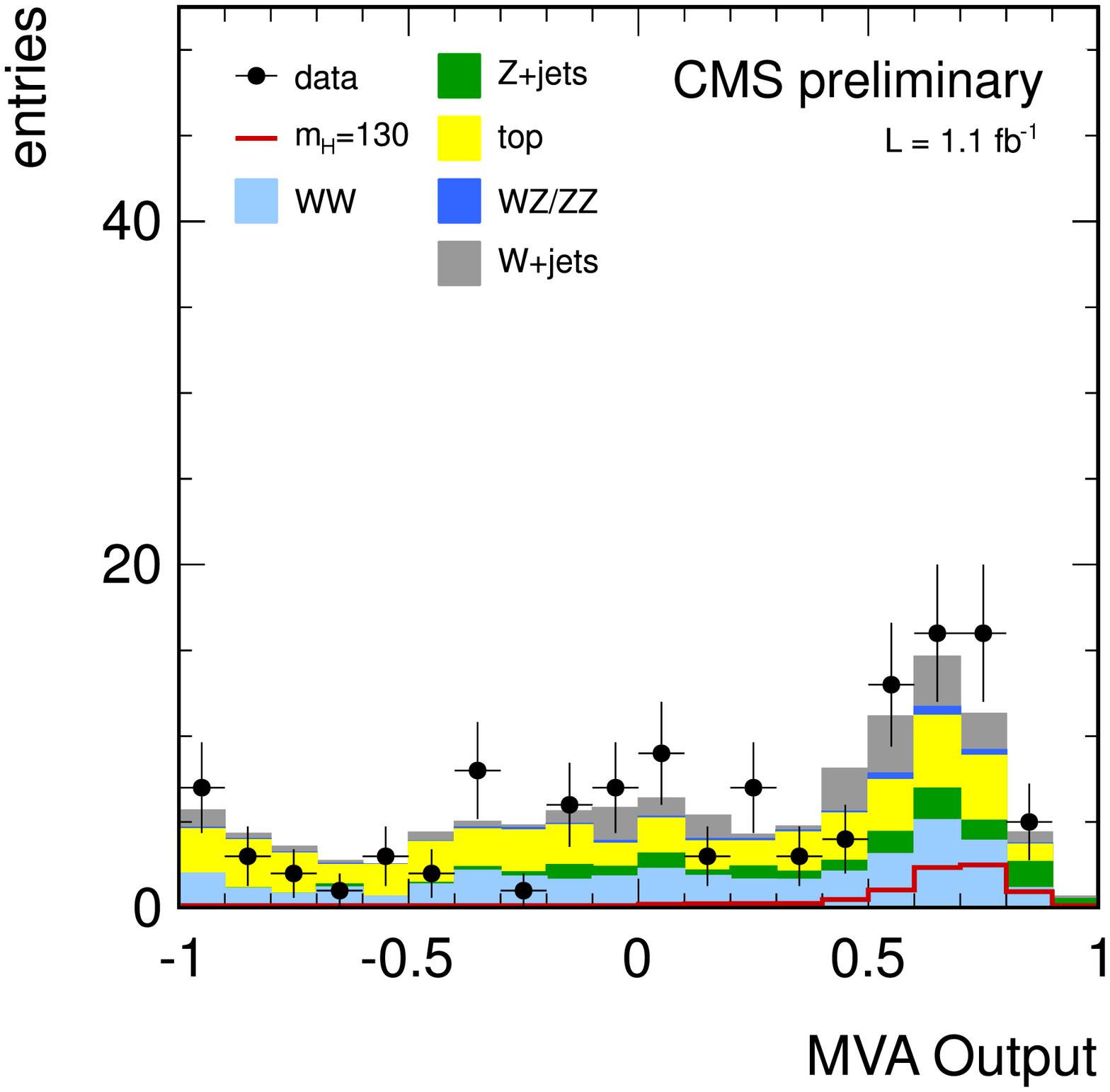}
\caption{BDT classifier outputs for Higgs signal and background events for 
$m_H=130\:$GeV$/c^2$ in the $0$-jet bin same flavor final state (top-left), $1$-jet bin same flavor 
final state (top-right), $0$-jet bin opposite flavor final state (bottom-left), and $1$-jet bin 
opposite flavor final state (bottom-right), after the $W^+W^-$ selection. The area marked as $WW$ 
corresponds to non-resonant $W^+W^-$ production.}
\label{fig:bdt_0j_1j}
\end{figure}

The kinematics of the Higgs boson decay depend on its mass, so the best performance is obtained by 
optimizing the selection requirements for each value of the mass. The values of the selections were 
chosen to minimize the upper limit on Higgs cross section, in the hypothesis of no signal. The same 
cut values are used in the $0$-jet and $1$-jet categories. In the multivariate approach, a boosted 
decision tree (BDT) algorithm~\cite{ref29} is used, trained for each Higgs boson mass hypothesis and
jet category. In addition to the $W^+W^-$ selection requirements, a loose cut on the upper value of 
$m_{\ell\ell}$ is applied to enhance the signal-to-background ratio. The multivariate technique uses
the following additional variables compared to the cut-based analysis: 
$\Delta R_{\ell\ell} \equiv \sqrt{\Delta\eta^2_{\ell\ell} + \Delta\phi^2_{\ell\ell}}$ between the 
leptons, with $\Delta\eta_{\ell\ell}$ being the $\eta$ difference between the leptons, which has 
similar properties as $\Delta\phi_{\ell\ell}$; the transverse mass of both 
lepton-$E_T^{\mbox{\scriptsize{miss}}}$ pairs; and finally, the lepton flavors. The training is 
performed using $H\rightarrow W^+W^-$ as signal and $W^+W^-$ continuum as background. The BDT 
outputs for $m_H=130\:$GeV$/c^2$ are shown in Figure~\ref{fig:bdt_0j_1j} for $0$-jet and $1$-jet 
events. The full distribution of the BDT output is used to compute confidence levels. To enhance the
sensitivity to Higgs signal, we use two categories: same-flavor dilepton ($ee/\mu\mu$) events and 
opposite flavor dilepton ($e\mu$) events.

\subsection{Strategy in the 2-jet Category}
The $2$-jet category is mainly sensitive to the VBF production mode. The cross section for this 
mode is roughly ten times smaller than the gluon-gluon fusion mode. However, the VBF signal can be 
extracted using simple selections, especially in the fully leptonic decay mode where the backgrounds
are expected to be relatively low.

The $H\rightarrow W^+W^-$ events from VBF production are characterized by a pair of energetic 
forward-backward jets and very little hadronic activity in the rest of the event. As a starting 
point, we select events that pass the preselection requiring two reconstructed jets with 
$p_T>30\:$GeV$/c$ and no other jets with $p_T>30\:$GeV$/c$. To reject the main background from top 
decays we apply two further improvements on the two jets $j_1$ and $j_2$: 
$|\Delta\eta(j_1-j_2)|>3.5$ and $m_{j_1j_2} > 450\:$GeV$/c^2$.

To improve the selection efficiency for the low mass Higgs signal we select events with the same 
$m_{\ell\ell}$ requirements as the $0$-jet and $1$-jet strategy. The VBF contribution to the total 
$H\rightarrow W^+W^-$ signal in the $2$-jet category is found to be about $85\%$ after the full 
event selection for all the mass hypotheses considered.

\section{Background Estimation}
Data driven methods are used as much as possible to estimate background contributions to the Higgs 
signal region. The top background and the $W+$jets and QCD multi-jets background are deduced from 
the measuring the efficiency of particular requirements designed to reject such events and applying 
the efficiency to a sample of rejected events. The Drell-Yan background in the dielectron and dimuon
final states and the non-resonant $W^+W^-$ background are estimated by scaling the expectation from 
simulation by a normalization factor between data and simulation in a signal-free control region.

Other backgrounds are estimated from simulation. The $W\gamma$ background estimate was cross-checked
in data using the events passing all selection requirements, except that the two leptons must have 
the same charge. This sample is dominated by $W$+jets and $W\gamma$ events. For $WZ$, $ZZ$ and 
$Z/\gamma^*\rightarrow\tau^+\tau^-$, good agreement between the data and the prediction
is found.

\subsection{W$+$jets and QCD multi-jets}
$W$+jets and QCD multi-jet events form a background to $W^+W^-$ production when jets are 
misidentified as leptons. The normalization and relevant kinematic distributions are estimated 
directly from data. A sample of loosely selected lepton candidates is extracted from events 
dominated by di-jet production. The probability for those candidates to be misidentified as a lepton
passing all lepton criteria, referred to as \emph{fake rate}, is calculated. The fake rate is 
applied to a sample of events selected with the analysis criteria except for one of the leptons for 
which the selection has been relaxed to the looser criteria and that fails the nominal selection.

\subsection{Top}
The top background is estimated from data by counting top-tagged events and applying the 
corresponding tagging efficiency. The tagging efficiency is measured in a data control sample with 
one counted jet, which is dominated by $t\bar{t}$ and $tW$ events. The top-tagging corresponds to 
the combination of soft-muon and $b$-jet tagging. 

\subsection{Drell-Yan}
An estimate of the residual $Z$ boson contribution to the $e^+e^-$ and $\mu^+\mu^-$ final states is 
obtained by extrapolating the observed number of events in data with dilepton invariant mass within 
$\pm15\:$GeV$/c^2$ of the nominal $Z$ mass. Before performing the extrapolation to the signal 
region, non-Drell-Yan processes contributing to the control sample must be subtracted. The 
contribution of $WZ$ and $ZZ$ in the $Z$ mass region, when leptons come from the same $Z$ boson, is 
subtracted as estimated from simulation. Other contributions, including $WW$, top, and $WZ$ and $ZZ$
when both leptons do not come from a $Z$ boson, are subtracted from an estimate using the 
$e^\pm\mu^\mp$ event yield in the $Z$ mass region, taking into account combinatorics and relative 
detection efficiencies for electrons and muons.

\subsection{WW}
The non-resonant $W^+W^-$ contribution in the $H\rightarrow W^+W^-$ low mass signal region, 
$m_H<200\:$GeV$/c^2$, can be estimated from data. This is done using events with a dilepton mass 
larger than $100\:$GeV$/c^2$, where there is a negligible contamination from the Higgs boson signal.
For larger Higgs boson masses there is a large overlap between the non-resonant $W^+W^-$ and Higgs 
boson signal. In these cases we estimate it from simulation.

\section{Results}
The yields after preselection and the expected background contributions are summarized in 
Table~\ref{tab:ww_yields}.

\begin{table}[h]
\begin{center}
\caption{Expected number of signal and background events from the data-driven methods for an 
integrated luminosity of $1.1\:$fb$^{-1}$ after applying the $W^+W^-$ selection requirements. 
Statistical and systematic uncertainties on the processes are reported. The 
$Z/\gamma^*\rightarrow\ell^+\ell^-$ process corresponds to the dimuon and dielectron final states. 
The $W^+W^-$ contribution corresponds to the estimated value from the simulation corrected for
lepton and trigger efficiencies.}
\begin{tabular}{|l|c|c|c|}
\hline 
\textbf{Process} & \textbf{0-jet} & \textbf{1-jet} & \textbf{2-jet} \\
\hline
$qq\rightarrow W^+W^-$             & $349.7\pm30.3$ & $101.4\pm9.3$  & $22.1\pm2.0$ \\
$gg\rightarrow W^+W^-$             & $17.2\pm1.6$   & $5.9\pm0.5$    & $1.1\pm0.1$ \\
$t\bar{t}+tW$                      & $63.8\pm15.9$  & $141.1\pm14.1$ & $99.3\pm9.9$ \\
$W+\gamma$                         & $8.7\pm1.7$    & $2.4\pm0.8$    & $1.1\pm0.5$ \\
$WZ+ZZ$                            & $8.5\pm0.9$    & $7.2\pm0.8$    & $1.5\pm0.2$ \\
$Z/\gamma^*\rightarrow\ell^+\ell^-$ & $12.2\pm5.3$   & $10.5\pm11.5$  & $19.2\pm13.5$ \\
$Z/\gamma^*\rightarrow\tau^+\tau^-$ & $1.6\pm0.4$    & $10.6\pm1.2$   & $3.9\pm0.7$ \\
$W+\mbox{jets}$                    & $106.9\pm38.9$ & $36.9\pm13.8$  & $16.4\pm6.4$ \\
\hline
all bkg. & $568.6\pm52.2$ & $316.0\pm24.7$ & $164.6\pm18.0$ \\
\hline
data & $626$ & $334$ & $175$ \\
\hline
\end{tabular}
\label{tab:ww_yields}
\end{center}
\end{table}

After applying the Higgs mass-dependent selections, no significant excesss is found with respect to 
the expected SM backgrounds, therefore upper limits are derived on the product of the Higgs boson 
production cross section and the $H\rightarrow W^+W^-$ branching fraction, 
$\sigma_H\times\mbox{BR(}H\rightarrow W^+W^-\rightarrow \ell\nu\ell'\nu'$), with respect to the SM 
expectation, ($\sigma^{95\%}/\sigma^{SM}$). The reported results come from a statistical method 
based on the hybrid Frequentist-Bayesian approach, known as $CL_S$~\cite{ref36}.

\begin{figure}[htb]
\centering
\includegraphics[width=80mm]{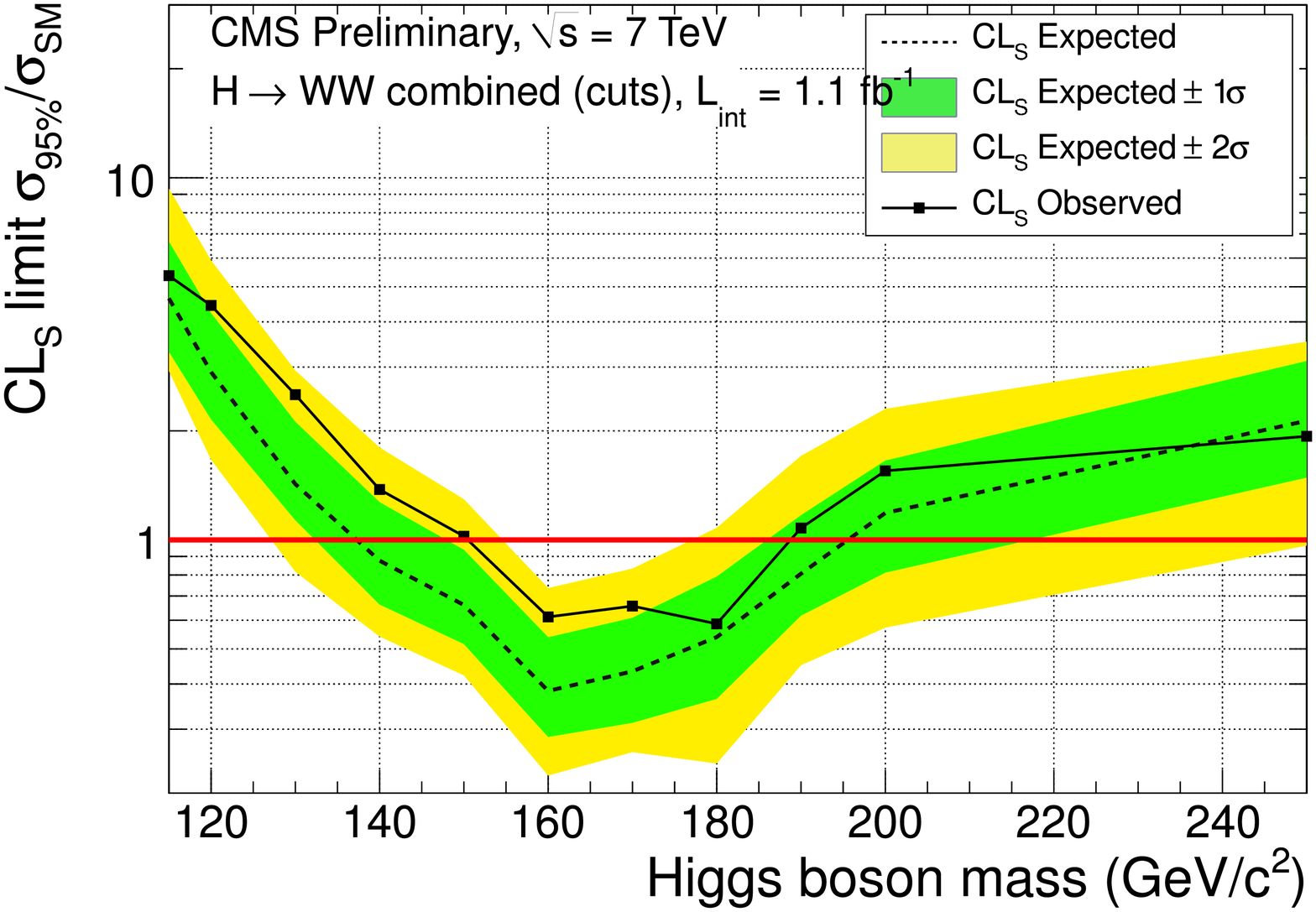}
\includegraphics[width=80mm]{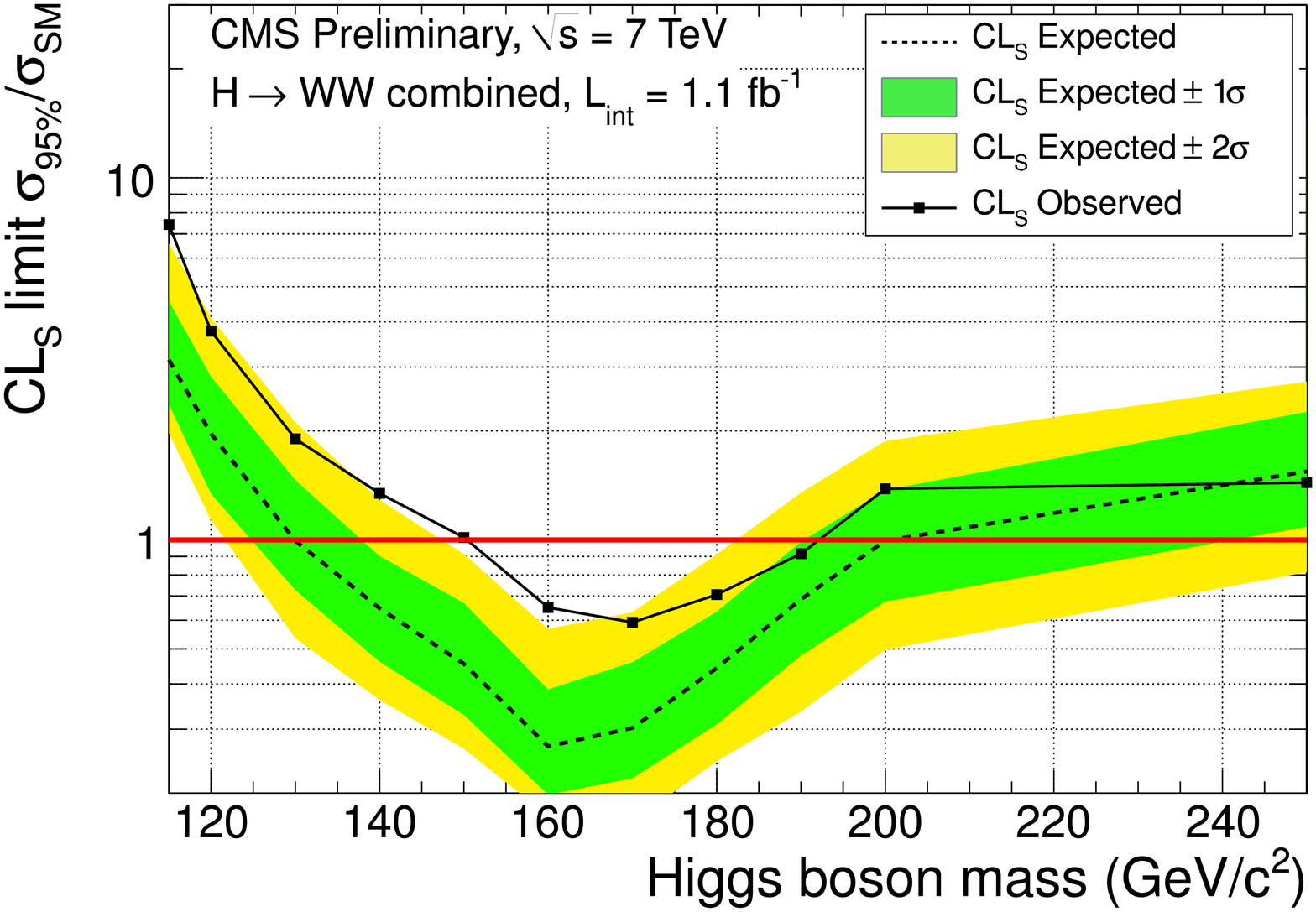}
\caption{Cut-based (left) and shape-based (right) modified frequentist 95\% C.L. expected (dotted 
line) and observed (points connected by solid line) upper limits on the cross section times 
branching ratio relative to the SM value.}
\label{fig:cls_hww_zoom}
\end{figure}

The $95\%$ observed and expected mean C.L. upper limits are shown in Figure~\ref{fig:cls_hww_zoom}.
Results are reported for both the cut-based and the BDT approach. The bands represent the $1\sigma$ 
and $2\sigma$ probability intervals around the expected limit. The a-posteriori probability 
intervals on the cross section are constrained by the a-priori minimal assumption that the signal 
and background cross sections are positive definite.

The multivariate analysis is chosen as the reference one because of better Higgs sensitivity.
The expected limits for any given mass hypothesis is better by about $30\%$ for the multivariate
analysis over the cut-based analysis, and the expected exclusion region is about $10\:$GeV$/c^2$
larger. We exclude the presence of a Higgs boson with a mass in $\left[150,\,193\right]\:$GeV$/c^2$ 
range at $95\%$ C.L. The observed exclusion range obtained with the cut-based approach is 
approximately the same. The observed (expected) upper limits are about 0.6 (0.3) times larger 
than the SM expectation for $m_H=160\:$GeV$/c^2$.

\section{Conclusion}
A search is performed for the Higgs boson decaying to $W^+W^-$ in pp collisions at 
$\sqrt{s}=7\:$TeV, using a data sample corresponding to an integrated luminosity of 
$1.1\:$fb$^{-1}$. No significant excess above the SM background expectation is found, and therefore
limits on the Higgs boson production cross section are derived, excluding the presence of a Higgs 
boson with a mass in $\left[150,\,193\right]\:$GeV$/c^2$ range at $95\%$ C.L.

\bigskip % extra skip inserted


\begin{thebibliography}{99}

\bibitem{ref1} K. Nakamura et al.(Particle Data Group), ``Review of particle phyics'', 
\emph{J. Phys.} {\bf G37} (2010) 075021. \verb=doi:10.1088/0954-3899/37/7A/075021.=

\bibitem{ref2} F. Englert and R. Brout,``Broken symmetries and the masses of gauge bosons'', 
\emph{Phys. Rev. Lett.} {\bf 13} (1964) 321. \verb=doi:10.1103/PhysRevLett.13.321.=

\bibitem{ref3} P.W. Higgs, ``Broken symmetry and the mass of gauge vector mesons'', 
\emph{Phys. Rev. Lett.} {\bf 13} (1964) 508. \verb=doi:10.1103/PhysRevLett.13.508.=

\bibitem{ref4} G. Guralnik, C.Hagan, and T. Kibble, ``Global Conservation Laws and Massless 
Particles'', \emph{Phys. Rev. Lett.} {\bf 13} (1964) 585-587. \verb=doi:10.1103/PhysRevLett.13.585.=

\bibitem{ref8} CMS Collaboration, ``Measurement of WW Production and Search for the Higgs Boson in 
pp Collsions at $\sqrt{s}=7\:$TeV'', \verb=arXiv:1102.5429.=

\bibitem{hww_pas_eps} CMS Collaboration, ``Search for the Higgs Boson in the Fully Leptonic 
$W^+W^-$ Final State'', \emph{CMS Physics Analysis Summary} {\bf CMS-PAG-HIG-11-003} (2011).

\bibitem{ref9} LHC Higgs Cross Section Working Group, S. Dittmaier, C. Mariotti et al., ``Handbook
of LHC Higgs Cross Sections: Inclusive Observables'', \verb=arXiv:1101.0593.=

\bibitem{ref20} CMS Collaboration, ``Particle Flow Event Reconstruction in CMS and Performance for 
Jets, Taus, and MET'', \emph{CMS Physics Analysis Summary} {\bf CMS-PAS-PFT-09-001} (2009).

\bibitem{ref21} CDF Collaboration, ``Measurement of the $W^+W^-$ production cross section and search
for anomalous $WW\gamma$ and $WWZ$ couplings in $p\bar{p}$ collisions at $\sqrt{s}=1.96\:$TeV'', 
\emph{Phys. Rev. Lett.} {\bf 104} (2010) 201801. \verb=doi:10.1103/PhysRevLett.104.201801.=

\bibitem{ref22} CMS Collaboration, ``Commissioning of the Particle-Flow reconstruction in 
Minimum-Bias and Jet Events from pp Collisions at 7 TeV'', \emph{CMS Physics Analysis Summary} 
{\bf CMS-PAS-PFT-10-002} (2010).

\bibitem{ref24} M. Cacciari, G.P. Salam, and G. Soyez, ``The anti-$k_t$ jet clustering algorithm'',
\emph{JHEP} {\bf 04} (2008) 063, \verb=arXiv:0802.1189.=

\bibitem{ref25} M. Cacciari and G.P. Salam, ``Dispelling the $N^3$ myth for the $k_t$ jet-finder'',
\emph{Phys. Lett.} {\bf B641} (2006) 57-61, \verb=arXiv:hep-ph/0512210.= 
\verb=doi:10.1016/j.physletb.2006.08.037.=

\bibitem{ref26} M. Cacciari and G.P. Salam, ``Pileup subtraction using jet areas'', 
\emph{Phys. Lett} {\bf B659} (2008) 119-126, \verb=arXiv:0707.1378.= 
\verb=doi:10.1016/j.physletb.2007.09.077.=

\bibitem{ref27} CMS Collaboration, ``Algorithms for b-jet identification in CMS'',
\emph{CMS Physics Analysis Summary} {\bf CMS-PAS-BTV-09-001} (2009).

\bibitem{ref28} CMS Collaboration, ``Commissioning of b-jet identification with pp collisions at 
$\sqrt{s}=7\:$TeV'', \emph{CMS Physics Analysis Summary} {\bf CMS-PAS-BTV-10-001} (2010).

\bibitem{ref29} A. Hoecker et al., ``TMVA - toolkit for multivariate data analysis'', 
\verb=arXiv:0703039.=

\bibitem{ref36} R.D. Cousins and V.L. Highland, ``Incorporating systematic uncertainties into an 
upper limit'', \emph{Nucl. Instrum. Meth.} {\bf A320} (1992) 331. Revised version. 
\verb=doi:10.1016/0168-9002(92)90794-5.=
\end{thebibliography}
\end{document}